# Nonlinear optical birefringence in $Li_2SO_4$-MgO-$P_2O_5$ amorphous system-influence of Cu ions

**A.Siva Sesha Reddy [a], A.V.Kityk [b], J.Jedryka [b,*], P.Rakus [b], A.Wojciechowski [b], A.Venkata Sekhar [a], V.Ravi Kumar [a], N.Veeraiah [a,*]**

[a] *Department of Physics, Acharya Nagarjuna University, Nagarjuna Nagar 522 510, India*
[b] *Faculty of Electrical Engineering, Czestochowa University of Technology, Armii Krajowej 17, PL-42-201 Czestochowa, Poland*

A B S T R A C T

Li₂SO₄-MgO-P₂O₅ glasses doped with varied contents of CuO (from 0- 1.0 mol%) were prepared. The optical birefringence induced by a High Q-2 femtosecond pulsed laser is explored as a function of copper oxide concentration in these glasses. The magnitude of the light-induced birefringence exhibited a gradual decrease with the increase of copper oxide content. The observed behavior is discussed in terms of polymerization of the glass network, the degree of which rises with concentration of $Cu^+$ ions occupying $T_d$ sites in the glass matrix. However, the samples containing low content of CuO are dominated by $Cu^{2+}$ ions that act as modifiers and induce structural imperfections by breaking of P-O-S bonds. Accordingly, the samples mixed with lower content of CuO are more favorable for inducing higher magnitudes of the optical birefringence. This conclusion is corroborated by results of electrical conductivity, positron annihilation and ultrasonic measurements.

## 1. Introduction

The investigations on nonlinear optical (NLO) properties in the amorphous materials attracted much interest during the past few decades [1-3]. Relevant investigations appear to be highly helpful when assessing an appropriateness of glass materials for optoelectronic applications, among which a number of 3D photonic devices, integrated optics and several NLO devices e.g. wide band optical amplifiers, high speed optically operated switches, power limiters [4,5]. The traditional methods viz. ion-exchange processes, light dissemination into a translucent substrate and also several other techniques, e.g. micro-fabrication and lithography that are being based on non-linear optical processes, are in general restricted to the construction of planar or lower-dimensional assemblies. However, when the glass materials are exposed to the laser pulses of short duration with appropriate intensity, a redistribution of the electronic charge density takes place facilitating the modification of internal microstructure of the material. Crucial role usually plays here is the multiphoton processes resulting in the non-linear absorption and/or light-induced optical anisotropy, i.e. the nonlinear optical birefringence. The light-induced birefringence is dependent on the laser wavelength, its energy, repetition rate and irradiation time. The internal structural disorder of the glass, color centers and excitons formed in the material due to irradiation with laser beam also play a predominant role in inducing the birefringence in the glasses. The nature and magnitude of the induced optical anisotropy and nonlinear absorption usually are characterized in terms of higher-rank tensor components of the complex nonlinear electrical susceptibility $\chi^{(n)} = \mathrm{Re}\chi^{(n)} + j\mathrm{Im}\chi^{(n)}$ ($n \geq 3$) describing variations both in the refractive index ($\mathrm{Re}\chi^{(n)}$, phase response) and in absorption coefficient ($\mathrm{Im}\chi^{(n)}$, amplitude response) that occur in the studied materials when exposed to powerful laser light.

So far, the majority of the studies available on optically stimulated NLO properties are restricted to SiO₂ based glass systems probably due to their broad applications as optical materials. Among various other possible glass systems that exhibit non-linear optical effects, alkali sulfophosphate glasses are of particular interest. In these glasses, the diffusion of sulphate ($SO_4^{2-}$) ions in to phosphate glasses is reported to be larger. However, in case, if phosphate glass network consists of metaphosphate structural groups, such dissolution is low and there is a possibility for the formation of thiophosphate ($SPO_7$)³⁻ groups with sulphate ions [6]. In other words, depending on the ratio of the alkali sulphate and P₂O₅ in the glass matrix, certain dynamic variations of associations and dissociations between SO₄ and PO₄ structural groups indeed take place. P–O–S interactions in this glass system have been particularly confirmed in nuclear magnetic resonance (NMR) studies [7]. This type of exchange interactions between sulfate and phosphate groups enables the dispersion of alkali ions into the glass network.

One should notice that the alkaline metal oxide MgO is admixed to the currently studied glass composition aiming to improve the corrosion resistance of $P_2O_5$ glass. Such modification forms polyhedrons with phosphate tetrahedrons enhancing thereby a durability of the glass matrix [8, 9]. The glasses containing the clusters of alkali/alkaline thiophosphate units admixed with different transition metal ions, on the other hand, are expected to exhibit higher magnitudes of non-linear optical susceptibilities improving thus their NLO conversion efficiency when exposed to high energetic laser beams [10].

Among various transition metal ions, Cu ions, that exist in $Cu^+$, $Cu^{2+}$ and occasionally in $Cu^{3+}$ states, are very interesting since they facilitate nonlinear optical anisotropy in the amorphous systems. In sulfophosphate glasses, the Cu ions are expected to be either in $Cu^+$ ($3d^{10}4s^1$) or $Cu^{2+}$ ($3d^{9}4s^1$) oxidation states. The ratio of these ions in the glass matrix depends on the content of both CuO as well as other constituents. Our recent dielectric, elastic and positron annihilation spectroscopy studies of copper oxide doped $Li_2SO_4$-MgO-$P_2O_5$ glass system [11-13] give a detailed characterization of the valence states of copper ions likewise their influence on structural variations occurring in the host glass matrix. To mention briefly, these investigations have suggested an increasing proportion of Cu ions in $Cu^+$ state that reside in $T_d$ sites in the glass matrix and alternate with $PO_4$ and $SO_4$ groups. One expects that such types of chemical bonding enhance the degree of polymerization occurring in the glass network. The elastic properties as well as positron annihilation spectroscopy studies indeed proved an increased rigidity of CuO-doped glasses with increasing of copper content. Such modifications in the glass matrix are predicted to induce an optical anisotropy when exposed to powerful femtosecond laser radiation.

The conventional glasses are optically isotropic and macroscopically centrosymmetric. However, the inversion symmetry may be broken under optical poling as has been demonstrated in pioneer work by Antonyuk et al [14] and confirmed later on by Balakirev et al [15] on commercial silica glasses. In relevant experiments interfering fundamental ($\omega$) and frequency-doubled ($2\omega$) laser waves produce in a glass media a weak static phase matched electric field wave $E_{dc} \propto \exp[i(q_{2\omega} - 2q_\omega)z]$ with spatial periodicity $L = 2\pi/K$ ($K = q_{2\omega} - 2q_\omega$). Optically poled samples are characterized thereby by the long-lived quadratic susceptibility being periodical across the sample and resulting in appearing of the grating.

The anisotropy in glass media, on the other hand, may be well induced also by polarized light exposing glass sample in a single pumping laser beam of fundamental frequency $\omega$. When light of high intensity propagates through a glass media this causes nonlinear Kerr effect [16]. Of particular interest are changes in the refractive index $n_i = n_0 + \delta n_i(I)$ in a material subjected to intensive light illumination of intensity $I$. Here $n_0$ is the linear refractive index, whereas the nonlinear contribution $\delta n_i$ is defined as

$$\delta n_i \propto \chi_{ijj}^{(3)}(\omega)\left\langle E_j^2(q,\omega)\right\rangle = \frac{1}{2}\chi_{ijj}^{(3)}(\omega)E_{0j}^2 \propto I, \qquad (1)$$

where indices $i$, $j$ and $k$ corresponds to Cartesian directions ($x$, $y$, $z$) coordinate system, $E_{0j}$ is the amplitude of light electric field, $q$ is the wavevector of the light wave, $\chi_{ijj}^{(3)}$ is the real part of the third order susceptibility representing a fourth-rank polar tensor (81 components). The number of independent components in isotropic media such as e.g. glasses or liquids is reduced just to two: $\chi_{1111}^{(3)} = \chi_{1221}^{(3)} + 2\chi_{1122}^{(3)}$ [17] and since $\chi_{2211}^{(3)} = \chi_{1122}^{(3)}$ one arrives with inequality: $|\chi_{1111}^{(3)}| \neq |\chi_{2211}^{(3)}|$. Accordingly, the polarized pumping light $E \parallel X$ results in the induced uniaxial anisotropy characterizing by the refractive indices: $n_1$, $n_2 = n_3$. Assuming that the wave vector of probing light $q \perp E$ it results to light-induced birefringence in that direction $\Delta n = n_1 - n_2 \alpha (\chi_{1111}^{(3)} -$

$\chi_{2211}^{(3)})E_{01}^2$. In experimental aspects it brings significant benefits taking into account high accuracy of the optical birefringence measurements. Dependent on a polarimetry technique used in birefringence measurements it is from 2 to 4 order of magnitude better compared to most accurate interferometric setups employed usually in the refractive index measurements. In fact, such technique, i.e. a combination of femtosecond pumping laser and a polarimetry setup with low power probing laser, has been applied in this work to explore nonlinear optical properties of CuO-doped $Li_2SO_4$-MgO-$P_2O_5$ glass material. One should emphasize, however, that the Kerr effect represents purely electronic nonlinearity, i.e. it is very fast with the time response in the femtosecond range. The light-induced birefringence in CuO-doped $Li_2SO_4$-MgO-$P_2O_5$ glasses exhibits, in contrast, slow relaxation behavior extended on a dozen of seconds. The nonlinear optical studies reported in this work evidently demonstrate that certain other nonlinear optical process(es) are present here in addition to the fast Kerr response. The observed behavior is discussed in terms of glass matrix polymerization mechanism and is analyzed in the light of recently reported experimental results on electrical, acoustic properties and positron annihilation studies.

## 2. Experimental

The following doped compositions of the glass (in mol%) have been chosen for the present study: $20Li_2SO_4$-$20MgO$-$(60-x)P_2O_5$: $xCuO$ ($x = 0$, 0.2, 0.4, 0.6, 0.8 and 1.0) referred hereafter as $C_0$, $C_2$, $C_4$, $C_6$, $C_8$ and $C_{10}$, respectively, i.e. relevant to CuO content. The particulars of procedure used for synthesis of the glass samples and the methods employed for characterization and the details of measurements followed for conductivity, acoustic properties and positron annihilation studies can be found in the Refs. [11-13].

The sketch of the laboratory nonlinear optical setup used for light-induced birefringence (LIB) measurements is shown in Fig.1. It combines the high accuracy photoelastic modulation polarimeter, equipped by the continuous-wave (CW) low power He-Ne laser ($\lambda$=633 nm, $P$=10 mW), and femtosecond pulsed Yb-based laser High Q-2 ($\lambda$=1045 ±5 nm, pulse duration ~ 250 fs, pulse repetition rate 63±0.6 MHz, spectral bandwidth FWHM 4-8 nm, peak radiant energy 40 nJ). The femtosecond laser play here role of strong pumping source (~2.5× $10^5$ W in pulse duration) inducing nonlinear birefringence. The low power CW He-Ne laser, being integrated into the polarimetry arm, is a probing monochromatic light source having no influence on the optical properties of media. The polarimetry arm (Fig. 1) consists of the photo-elastic modulator (PEM-90, Hinds Instr.) which along with a sample is placed between the crossed polarizer (P) and analyser (A). In such polarimetry scheme the photoelastic modulator retardation ($\Omega/2\pi = 50~kHz$) results to modulated light intensity, detected by the photo-detector (PD), and subsequently analysed by the pair of lock-in amplifiers (SR-830, Stanford Research) extracting the signal amplitudes of the first ($I_\Omega$) and second ($I_{2\Omega}$) harmonics of the modulated light. The light-induced optical retardation $\delta_x$ and LIB $\delta(\Delta n)$ is then defined as:

$$\delta_x = \tan^{-1}\left(\frac{I_\Omega}{I_{2\Omega}}\frac{J_2(\delta_0)}{J_1(\delta_0)}\right) ~ ; ~ \delta(\Delta n) = \frac{\lambda\delta_x}{360^\circ d}, \qquad (2)$$

where $J_1(\delta_0)$ and $J_2(\delta_0)$ are the values of Bessel function corresponding to the PEM amplitude of the modulated retardation ($\delta_0$ =0.383$\lambda$), $d$ is the sample thickness.

More precisely $\delta_0$ is the amplitude of phase delay modulation or the amplitude of retardation modulation that have the same meaning. One may express them wavelength units (e.g. nm) or angular units (e.g. degrees). Optical retardation of 1$\lambda$ is equivalent to 360°, thus 0.383$\lambda$ is equivalent to about 137.9°. The acquired data were transferred via GPIB to PC for their real-time processing and subsequent storing. The precision in polarimetry experiments was about 5× $10^{-3}$ deg. what is equivalent to the accuracy in the $\delta(\Delta n)$ measurements ~$10^{-8}$ for samples of ~ 1 mm thick.

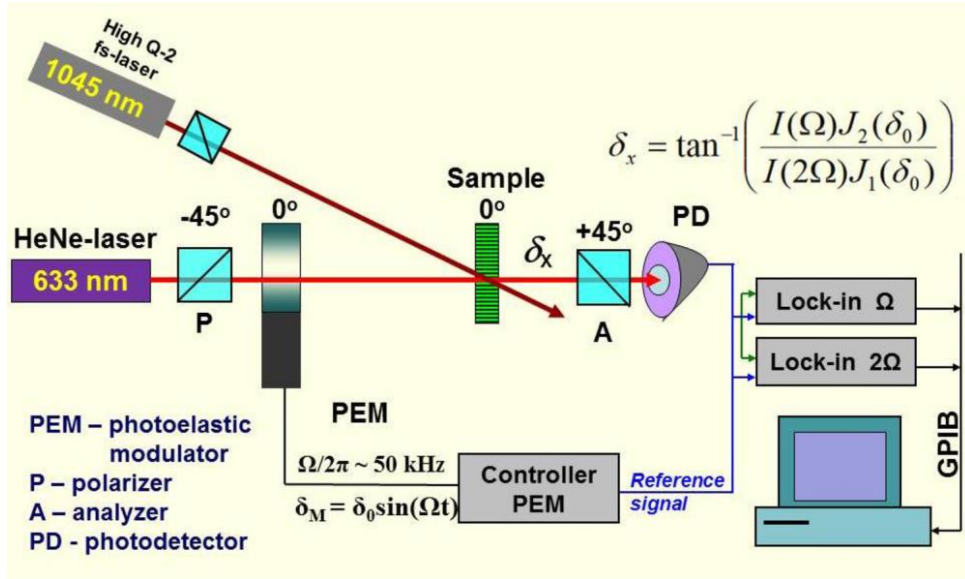

**Fig. 1.** A sketch of the NLO laboratory setup used for measurements of the light- induced birefringence.

## 3. Results and discussion

Fig. 2(a) shows the temporal behavior of optical retardation for CuO-doped glass sample $C_{10}$, as example, recorded at subsequent time intervals when the pumping laser light, illuminating the sample, alternatingly switches between states OFF and ON. Right in the moments of switching OFF/ON, or vice versa, one observes abrupt almost jump-like changes of the optical retardation which then reveals a long-time relaxation evolution extending on a dozen of second with a saturated character. The changes of the birefringence in subsequent OFF/ON intervals are somewhat different. The glass samples reveal also residual mechanical strain birefringence ($10^{-5}$ - $10^{-6}$) having local and random character (different magnitude and anisotropy directions) resulting usually during the preparation of the glass. Strain birefringence means that the materials exhibit optical anisotropy. Normally, in an ideal case amorphous materials, such as glasses, are optically isotropic having no birefringence. In reality, however, they usually exhibit small optical anisotropy originating during the preparation. Non-homogeneous temperature (temperature gradients) during cooling of melt and subsequent solidification result in static residual strains being the reason for optical anisotropy called usually as strain birefringence. Polarimetry techniques reveal it immediately analyzing the measured phase retardation δ. Strain birefringence then is defined as $\Delta n = \delta \lambda/(360^0 d)$, where d is the sample thickness. Indeed, $10^{-5}$ - $10^{-6}$ is small anisotropy. For comparison, weakly anisotropic crystal materials exhibit birefringence typically in the range $10^{-3}$ - $10^{-4}$, more-stronger like e.g. calcite or large number of liquid crystals reveal the birefringence of the order of $10^{-1}$.

For this reason we refer in our further analysis to the light-induced changes in the optical anisotropy ignoring strain birefringence. It means that we ignore the static (constant) contribution of residual birefringence and pay attention on temporal changes of birefringence induced by light illumination. In other words, static onset birefringence originating from stress birefringence is out of our interest.

Moreover, to give a proper quantitative characterization of the optical anisotropy, particularly to level the thickness factor, we present in Fig. 2(b) the temporal evolution of the optical birefringence recalculated directly from the measured optical retardation (see eq. (2)).

The powerful femtosecond laser radiation evidently induces the optical birefringence temporal behavior of which evidently appears in different time scales. The glasses, being macroscopically centrosymmetric, may exhibit only quadratic electro-optical properties like e.g. electro-optic Kerr effect. Assuming that we deal with electronic response one may call it as nonlinear optical Kerr effect [18]. This effect is expected to be very fast with rates appearing somewhere in THz region. Although a jump-like temporal behavior right away at ON/OFF switching may be attributed to the Kerr effect, further slow relaxation, extending on a dozen of seconds, evidently indicates on a presence of some other NLO mechanism. Extremely slow relaxation most likely may be attributed to a varied degree of disorder in the glass matrix due to the variation in the content of CuO.

To characterize light induced birefringence (LIB) changes we took difference between the saturated values of $\Delta n(t)$ when light is ON and OFF. Saturated values, in other words, mean that they were extracted somewhere near the ends of subsequent circles OFF-ON-OFF-ON-…. The variation of $\Delta n$ with the concentration of CuO is shown in Fig. 2(c). It may be noted here that we have also recorded LIB for the undoped (CuO) glasses; the samples have exhibited insignificant $\Delta n$ (<1.0× $10^{-5}$).

In addition thermal effects, indeed are possible and presumably are presented. Relevant thermal mechanism is related with temperature gradient effect which only may lead to induced thermal strains and relevant optical anisotropy. Unfortunately its quantitative characterization is not trivial. One may mention, however, that light absorption on Cu$^-$ ions and subsequent its conversion to heat energy seems to be not the case taking into account that photoinduced birefringence (see Fig. 2 (c)) decreases with rising of Cu-concentration. In other words, photorefractive mechanism in our opinion is here dominating.

Our earlier measurements (optical absorption, XPS [11-13]) indicate that the concentration of Cu$^+$ ions, occupying tetrahedral sites in the glass network, increases with the copper oxide content. The electrical conductivity measurements demonstrate a decaying trend with the rise of CuO concentration (Fig. 3(a)). The copper ions are predicted to exist in Cu$^{2+}$ state and act as modifiers and cause to increase the conductivity. However, during the preparation of the glass there is a possibility for some of these ions to reduce in to Cu+ state (proved by XPS studies, Refs. 11-13]. The Cu$^+$ ions (unlike Cu$^{2+}$ ions) do participate in the glass network with tetrahedral units and alternate with PO4 units. Such augmentation of the glass network is an hindrance for the migration of charge carriers that contribute to the conductivity. Hence, if there is a higher concentration of Cu+ ions, there will be a decrease in the conductivity.

The free volume fraction entrenched in the glass network, as evaluated by positron annihilation decay profiles measured with $^{22}Na$ isotope 0.1MBq, is found to be decreased with the CuO content (Fig. 3(b)). The ultrasonic velocity in these samples measured at frequency of 4 MHz exhibits an increasing trend on the CuO content (Fig.3(c)). It may be noted here that these figures were reproduced from the Refs [11-13] for the sake of reference.

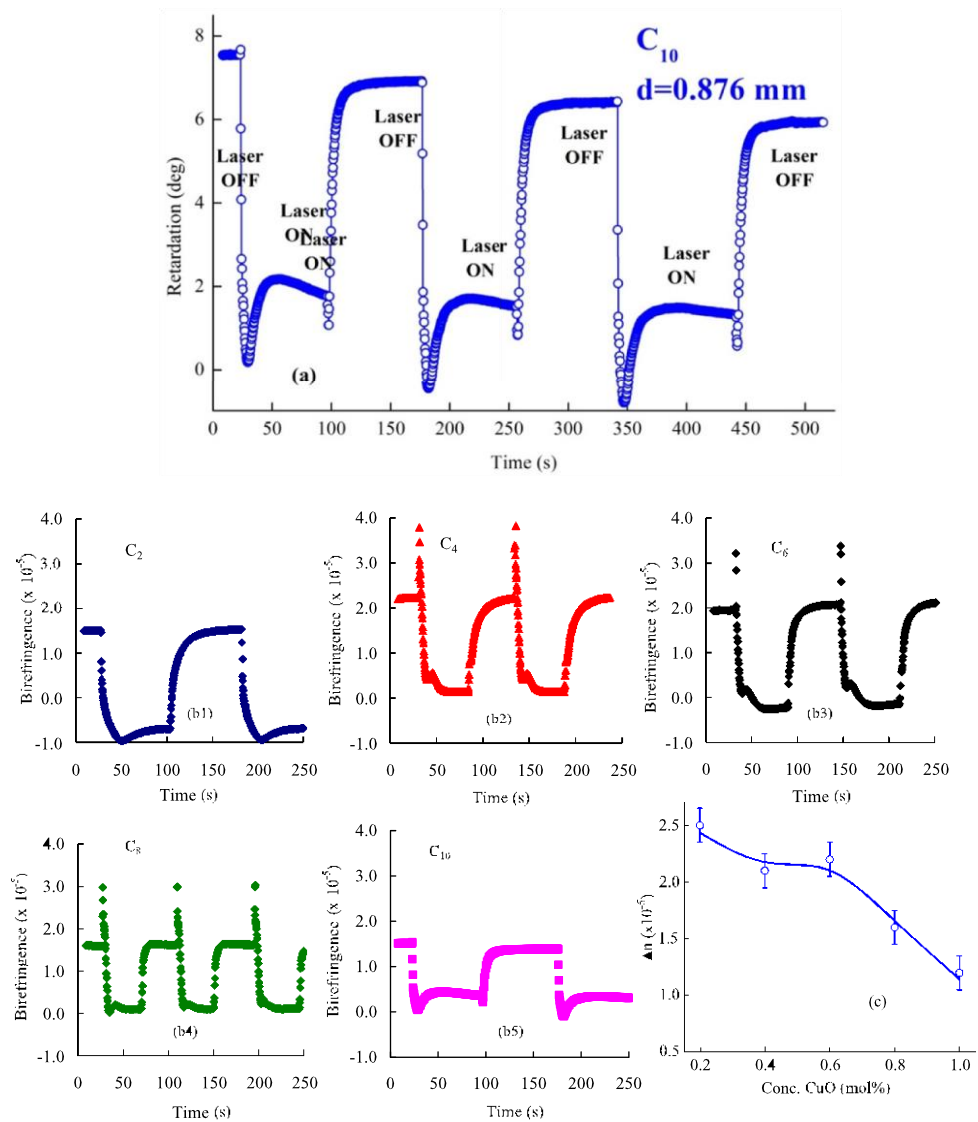

**Fig. 2.** Section (a): Retardation vs time recorded at subsequent temporal intervals when the pumping laser light, illuminating the sample C₁₀, alternately switches between states OFF and ON. Sections (b1)-(b5): Relevant changes of the optical birefringence induced in CuO-doped Li₂SO₄-MgO-P₂O₅ glasses under alternately switching pumping laser light. Section (c): The light-induced birefringence vs the copper oxide content.

Taking together, the results of these studies suggest an increased degree of polymerization of the glass network with the rise of CuO content. Such rigidity results in a decrease of the optical birefringence changes due to the phonon losses as has been noticed above. It may be noted here that in the samples containing low content of CuO, major portion of copper ions exists in Cu²⁺ state and act as modifiers defragmenting the glass network internally due to breaking of P-O-S bonds [19, 20]. Hence, the glass samples doped with low content of CuO are characterized by larger light-induced optical anisotropy and, accordingly, by lager magnitudes of LIB. Such materials appear to be therefore useful in optical systems that include telecom components (optical circulators and interleavers) likewise other applications such as e.g. brightness enhancers in illumination systems etc.

## 4. Conclusion

Li₂SO₄-MgO-P₂O₅ glasses mixed with various concentrations of copper oxide were synthesized. Spectroscopic, electrical conductivity,

positron annihilation and ultrasonic studies reported in detail in our recent works evidently proved that there is a rising degree of structural polymerization with increase of CuO content. It has been ascribed to the increasing concentration of Cu⁺ ions that participate in the glass network in tetrahedral sites and augment with PO₄ and SO₄ structural groups. The light-induced birefringence versus the content of copper oxide exhibited a decaying trend. Such decrease is attributed to the increased phonon losses in the highly polymerized glass matrix.

The other possible reasons for inducing LIB like residual mechanical strain birefringence are discussed but their contribution is predicted to be negligible. The decrease of the magnitude of LIB, which correlates with rising CuO concentration, is explained by the increased degree of polymerization of the glass matrix due to the rising of Cu⁺ ions concentration occupying $T_d$ sites in the glass matrix. It may be stressed here, that in the samples containing low content of CuO, most of copper ions appear to exist in Cu²⁺ state act as modifiers and de-augment thereby the glass network by breaking P-O-S bonds. Accordingly, the glasses mixed with lower CuO content are expected to be more favorable for inducing higher magnitudes of the optical birefringence.

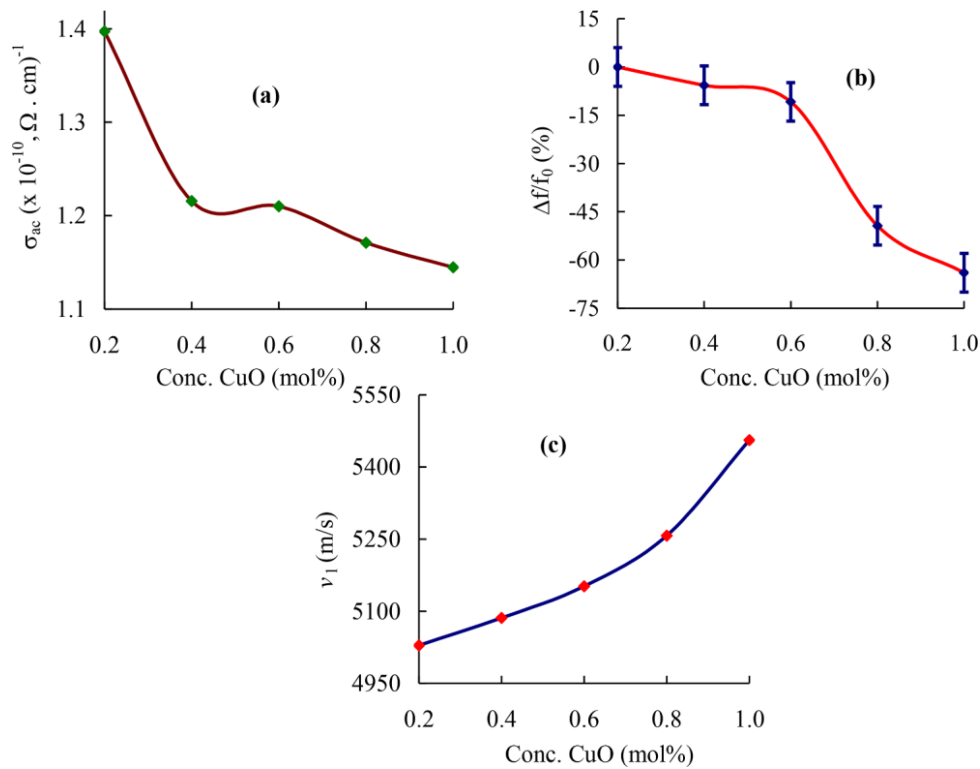

**Fig. 3.** Variations of the ac conductivity ($f = 300$ Hz) measured at $T = 100$ °C (Sec. (a)), the free volume fraction trapped in the glass matrix as evaluated by positron annihilation spectroscopy with 0.1 MBq $^{22}$Na isotope (Sec.(b)) and the longitudinal ultrasound velocity ($f = 4$ MHz) measured at ambient temperature (Sec.(c)) vs the CuO concentration in $Li_2SO_4$-MgO-$P_2O_5$ glasses.

## Credit Author statement

All the Authors have equally contributed to this manuscript.

## Acknowledgement

The presented results are part of a project that has received funding from the European Union Horizon 2020 research and innovation programme under the Marie Sklodowska-Curie grant agreement no. 778156. Support from resources for science in years 2018–2022 granted for the realization of international co-financed project Nr W13/H2020/2018 (Dec. MNiSW 3871/H2020/2018/2) is also acknowledged. One of the authors N. Veeraiah, wishes to thank UGC-New Delhi, for sanctioning BSR-Faculty Fellowship, to carry out this work..

## References

[1] K. Ogawa, T. Honma, T. Komatsu, Birefringence imaging and orientation of laser patterned β-BaB$_2$O$_4$ crystals with bending and curved shapes in glass, J. Solid State Chem. 207 (2013) 6–12.

[2] Zhen-Kai Fan, Shu-Guang Li, Wan Zhang, Guo-Wen An, Ya-Jie Bao, Analysis of the polarization beam splitter in two communication bands based on ultrahigh birefringence dual-core tellurite glass photonic crystal fiber, Opt. Commun. 333 (2014) 26–31.

[3] T.H. Pei, Z. Zhang, Y. Zhang, The high-birefringence asymmetric SF57 glass micro-structured optical fiber at 1060.0 μm, Opt. Fiber Technol. 36 (2017) 265–270.

[4] D. Ehrt, T. Kittel, M. Will, S. Nolte, A. Tunnermann, Femtosecond-laser-writing in various glasses, J. Non-Cryst. Solids 345-346 (2004) 332–337.

[5] K. Divakara Rao, K.K. Sharma, Dispersion of the induced optical nonlinearity in Rhodamine 6G doped boric acid glass, Opt. Commun. 119 (1995) 132–138.

[6] M. Ganguli, K.J. Rao, Studies of ternary Li$_2$SO$_4$–Li$_2$O–P$_2$O$_5$ glasses, J. Non-Cryst. Solids 243 (1999) 251–267.

[7] N. Hemono, F. Munoz, Dissolution of SO$_3$ within a lithium phosphate glass network and structure-property relationships, Phys. Chem. Glasses Eur. J. Glass Sci. Technol. B 51 (2010) 121–126.

[8] Xin Wang, Franscico Munoz, Dongbing He, Yajun Ding, Lili Hu, Effects of SiO$_2$ on properties and structures of neodymium doped P$_2$O$_5$–Al$_2$O$_3$–Li$_2$O–MgO–Sb$_2$O$_3$ glasses, J. Alloys Compd. 729 (2017) 1038–1045.

[9] M.A. Karakassides, A. Saranti, I. Koutselas, Preparation and structural study of binary phosphate glasses with high calcium and/or magnesium content, J. Non-Cryst. Solids 347 (2004) 69–79.

[10] He-Gen Zheng, Yu-Xiao Wang, Mei-Hua Ma, Ya Cai, Anwar Usman, Hoong-Kun Fun, Ying-Lin Song, Xin-Quan Xin, Crystal structures and nonlinear optical properties of new clusters [MOS$_3$Cu$_3$(PPh$_3$)$_3${S$_2$P(OCH$_2$Ph)$_2$}] (M/Mo, W), Inorg. Chim. Acta 351 (2003) 63–68.

[11] A. Venkata Sekhar, L. Pavić, A. Moguš-Milanković, N. Purnachand, A. Siva Sesha Reddy, G. Naga Raju, N. Veeraiah, Dielectric characteristics, dipolar relaxation dynamics and ac conductivity of CuO added lithium sulpho-phosphate glass system, J. Non-Cryst. Solids 543 (2020), 120157.

[12] A. Venkata Sekhar, M. Kostrzewa, Valluri Ravi Kumar, A. Ingram, A. Siva Sesha Reddy, G. Naga Raju, V. Ravi Kumar, N. Veeraiah, Estimation of concentration of nano-sized voids ingrained in CuO doped lithium sulphophosphate amorphous system using positron annihilation spectroscopy, Opt. Mater. 109 (2020), 110314.

[13] A. VenkataSekhar, A.V. Kityk, J. Jedryka, P. Rakus, A. Wojciechowski, A. Siva Sesha Reddy, G. Naga Raju, N. Veeraiah, Investigations on the influence CuO doping on elastic properties of Li$_2$SO$_4$–MgO–P$_2$O$_5$ glass system by means of acoustic wave propagation, Solid State Commun. 330 (2021), 114270.

[14] B.P. Antonyuk, N.N. Novikova, N.V. Didenko, O.A. Aktsipetrov, All optical poling and second harmonic generation in glasses: theory and experiment, Phys. Lett. 287 (2001) 161–168.

[15] M.K. Balakirev, I.V. Kityk, V.A. Smirnov, L.I. Vostrikova, I. Ebothe, Anisotropy of the optical poling of glass, Phys. Rev. A 67 (2003), 023806.

[16] M. Sheik-Bahae, D.J. Hagan, E.W. Van Stryland, Dispersion and bandgap scaling of the electronic Kerr Effect in solids associated with two-photon absorption, Phys. Rev. Lett. 65 (1990) 96–99.

[17] M.J. Weber, D. Milam, W.L. Smith, Nonlinear refractive index of glasses and crystals, Opt. Eng. 17 (1978) 463–469.

[18] D. Pourmostafa, H. Tajalli, A. Vahedi, K. Milanchianc, Electro-optical Kerr effect of 6CHBT liquid crystal doped with MgO nanoparticles in different concentration, Opt. Mater. 107 (2020), 110061.

[19] L. Pavic, A. Mogus-Milankovic, P. Raghava Rao, A. Santic, V. Ravi Kumar, N. Veeraiah, Effect of modifier ion on electrical, dielectric and spectroscopic properties of Fe$_2$O$_3$ doped Na$_2$O–MoO$_3$-MO-P$_2$O$_5$glass system, J. Alloys Compd. 604 (2014) 352–362.

[20] T. Srikumar, I.V. Kityk, Ch.Srinivasa Rao, Y. Gandhi, M. Piasecki, P. Bragiel, V. Ravi Kumar, N. Veeraiah, Photostimulated optical effects and some related features of CuO mixed Li$_2$O–Nb$_2$O$_5$–ZrO$_2$–SiO$_2$ glass ceramics, Ceram. Int. 37 (2011) 2763–2779.